\documentclass[aps,jcp,twocolumn,superscriptaddress,floatfix,10pt]{revtex4}
\usepackage{comment}
\usepackage{graphicx}
\usepackage{amsmath}
\usepackage{color}
\usepackage{dcolumn}
\usepackage{soul}
\usepackage{comment}
\usepackage{appendix}
\usepackage{hyperref}
\usepackage[capitalize]{cleveref}
\usepackage[thinc]{esdiff}

\renewcommand{\arraystretch}{0}

\begin{document}
	
	\title{Thermal conductivity of bottle--brush polymers}
	
	\author{Manoj Kumar Maurya}
	\affiliation{Department of Mechanical Engineering, Indian Institute of Technology Kanpur, Kanpur UP 208016 India}
	\author{Tobias Laschuetza}
	\affiliation{Institute of Mechanics, Karlsruhe Institute of Technology (KIT), Otto--Ammann--Platz 9, 76131 Karlsruhe, Germany}
	\affiliation{Quantum Matter Institute, University of British Columbia, Vancouver BC V6T 1Z4, Canada}
	\author{Manjesh Kumar Singh}
	\email{manjesh@iitk.ac.in}
	\affiliation{Department of Mechanical Engineering, Indian Institute of Technology Kanpur, Kanpur UP 208016 India}
	\author{Debashish Mukherji}
	\email{debashish.mukherji@ubc.ca}
	\affiliation{Quantum Matter Institute, University of British Columbia, Vancouver BC V6T 1Z4, Canada}
	
	\begin{abstract}
		Using molecular dynamics (MD) simulations of a generic model, we investigate heat propagation in bottle--brush polymers (BBP). 
		An architecture is referred to as a BBP when a linear (backbone) polymer is grafted with the side chains of different length $N_{\rm s}$ and 
		grafting density $\rho_{\rm g}$, which control the bending stiffness of a backbone. 
		Investigating $\kappa-$behavior in BBP is of particular interest due to two competing mechanics: increased backbone stiffness, via $N_{\rm s}$ and $\rho_{\rm g}$,
		increases the thermal transport coefficient $\kappa$, while the presence of side chains provides additional pathways for heat leakage. 
		We show how a delicate competition between these two effects controls $\kappa$. These results reveal that
		going from a weakly grafting ($\rho_{\rm g} < 1$) to a highly grafting ($\rho_{\rm g} \ge 1$) regime,
		$\kappa$ changes non--monotonically that is independent of $N_{\rm s}$. The effect of side chain mass 
		on $\kappa$ and heat flow in the BBP melts are also discussed. 
		
	\end{abstract}
	
	\maketitle
	
	\section{Introduction}
	\label{sec:intro}
	
	Understanding the structure--property relationship in polymers is at the onset of many
	developments in designing advanced functional materials for suitable applications~\cite{PolRevTT14,Mueller20PPS,Mukherji20AR,Keblinski20,PolRev2020Jie,nancy22rev}. 
	This is particularly because polymers are an important class of soft matter, where a delicate entropy--energy
	balance dictate their physical properties~\cite{PolRevTT14,Mueller20PPS,Mukherji20AR}. While the properties 
	of most commonly known polymers are governed by weak van der Waals (vdW) interactions, 
	the strength of which is of the order of $k_{\rm B}T$ under ambient temperature $T = 300$ K
	and $k_{\rm B}$ being the Boltzmann constant, the recent interest has been devoted to bio--compatible 
	hydrogen bonded (H--bond) polymers having a relatively stronger interaction strength of about 4--8 $k_{\rm B}T$~\cite{desiraju02,Pipe15NMat,Cahill16Mac,weil20jacs,DMKKpolRev23,review23}.
	
	The use of different polymeric materials ranges from commodity items~\cite{halek1988relationship,jain2011biodegradable,maier2001polymers,Mukherji19PRM},
	complex electronic packaging~\cite{Pipe15NMat}, thermoelectrics~\cite{pedot16,shi2017tuning,tripathi2020optimization}, organic solar cells~\cite{nancy22rev,smith2016high}, 
	and/or defense purposes~\cite{mcaninch2013characterization,elder2016nanovoid}, where they are often exposed to a variety of environmental conditions.
	These include, but are not limited to, temperature $T$, pressure $p$, and solvent conditions.
	In this context, one of the most intriguing properties of amorphous polymers is their 
	ability to conduct the heat current, as quantified by the thermal transport coefficient $\kappa$~\cite{PolRevTT14,Keblinski20,PolRev2020Jie}. 
	Here, $\kappa \propto c v_{\rm g}\Lambda$ with $c$, $v_{\rm g}$, and $\Lambda$ being the heat capacity, the group velocity, and the mean free path, 
	respectively. Note that $v_{\rm g}$ is related to the material stiffness. 
	In an amorphous polymer, where an individual polymer chain follows random walk statistics~\cite{DesBook,DoiBook,DGbook}), 
	$\Lambda \to 0$ and thus leads to a smaller $\kappa$. More specifically, most vdW--based systems usually 
	have $\kappa \to 0.1-0.2$ W/Km~\cite{Cahill16Mac} and can reach as high as 0.4 W/Km for H--bonded systems~\cite{Pipe15NMat,Cahill16Mac}.
	
	Even when the bulk $\kappa$ of amorphous polymers is rather small, at the monomer level they have different pathways for 
	energy transfer, i.e., between two bonded monomers and between a monomer and its non--bonded neighbors. 
	The energy transfer rate between two bonded monomers is over two orders of magnitude faster than between the
	non--bonded monomers~\cite{MM21acsn,MM21mac,wu22cms}. This behavior is expected because $\kappa$ is directly related
	to the material stiffness~\cite{Cahill90PRB}. For example, a carbon--carbon (C--C) bond has an elastic modulus of $ \simeq 250$ GPa~\cite{crist1996molecular}, 
	while the vdW and H--bonded systems have 1--5 GPa~\cite{Cahill16Mac}. 
	Note that we draw an analogy with a C--C bond because they constitute the backbone of most known commodity polymers. 
	
	The increased bonded energy transfer rates are shown to play a key role in increasing $\kappa$. Typical examples are
	the chain oriented systems of polymer fibers~\cite{GangChenNanoNature} and polymer brushes~\cite{bhardwaj2021thermal}, 
	attaining $\kappa \geq 100$ W/Km. Here, an extended chain configuration can be approximated as
	a quasi one--dimensional crystal due to the periodic arrangement of monomers and thus $\kappa$ is dictated by the phonon propagation
	along the direction of orientation. Phonon scattering occurs when there is a kink along the chain backbone~\cite{GangChenJAP} and/or 
	if there is an added pathway (such as a side chain) for heat leakage~\cite{kappaside18mat}, thus
	leads to a significant reduction in $\kappa$.
	
	Side chains are often added in a variety of important polymeric architectures to control their physical properties. 
	A typical class of experimentally relevant systems is conjugated polymers, 
	such as P3HT~\cite{nancy22rev,smith2016high} and super yellow~\cite{superyellow}. 
	The backbones of these systems are extremely stiff, hence they remain insoluble in most common solvents (due to the obvious 
	entropic effects). Short alkane chains are then attached to the backbones to improve their solubility 
	and thus improving solution processing. As a consequence, when these systems self--assemble for a device
	application, such as organic solar cell, alkane chains play a key role. For example, when a device is used under 
	the high $T$ conditions, one grand challenge is to remove the excess heat for its better performance. Typically this is done via adding 
	high $\kappa$ fillers~\cite{smith2016high} that often compromise the basic properties of the bare background material.
	In this context, it is particularly important to understand the heat propagation in the bare samples and, if possible,
	to provide a microscopic picture for the tunability in $\kappa$. 
	
	Another interesting system is the ``so called" bottle--brush polymers (BBP)~\cite{Sergei_1,binderJCP2011}.
	These are of potential interest because they have the potential to be used as one--dimensional organic nanocrystals~\cite{nanotube2016},
	low friction materials~\cite{lubrication}, solvent--free and supersoft networks~\cite{Sergei_2}, have 
	extraordinary elastic properties~\cite{Sergei_2}, and pressure--sensitive adhesives~\cite{Sergei_4}.
	Another important feature of BBP is that their
	backbone stiffness, as measured in the unit of Kuhn length $\ell_{\rm k}$, can be tuned by changing the 
	side chain length $N_{\rm s}$ and the grafting density $\rho_{\rm g}$. Here, $\ell_{\rm k} \propto N_{\rm s}^{\alpha}$ with $\alpha \simeq 0.5-1.0$
	for $N_{\rm s} \leq 40$~\cite{binderJCP2011}. One of the earliest studies also predicted $\alpha = 15/8$ in the asymptotic limit~\cite{marquesThesis1989}.
	
	The discussions presented above clearly show that a better microscopic picture is needed that provides a possible route
	to tune $\kappa$ in the branched systems via macromolecular engineering. 
	In this context, to the best of our knowledge, simulation studies dealing with $\kappa$ in such systems are rather limited, 
	except for one case study using all--atom simulations of backbone polynorbornene (PNB) grafted with polystyrene (PS) side chains~\cite{kappaHaoMaa}. 
	Motivated by this need, we investigate the influence of side chains on the $\kappa-$behavior in the branched polymers
	using a generic model. 
	For this purpose, we simulate a model system consisting of a set of BBPs with varying $N_{\rm s}$ and $\rho_{\rm g}$.
	We decouple various effects on $\kappa$ and show how a delicate balance between different system parameters control $\kappa$ in BBPs.
	{We note in passing that since the significance of such brush--like polymers range over a wide variety of polymer
		chemistry and thus investigating them at the all--atom level has its obvious limitations. Therefore, a generic model
		is more suitable, where a broad range of polymers can be represented within one generic framework, 
		while ignoring the specific chemical details that often only contribute to a mere numerical pre--factor.} 
	
	The remainder of the paper is organized under the following sections: method and model used for this study, results and discussions, and finally the conclusions are drawn.
	
	\section{Materials, model, and method}
	\label{sec:mm}
	
	In this study we investigate two systems: namely a set of tethered BBPs and a set of BBP melts. 
	In both cases, a BBP consists of a backbone of length $N_{\rm b}$ grafted with the side chains of different $N_{\rm s}$ and 
	$\rho_{\rm g}$, where $\rho_{\rm g} = N_{\rm g}/\lambda$. $N_{\rm g}$ is the number of chains grafted per backbone monomer 
	and $\lambda$ is the monomer distance along the backbone for grafting. 
	
	\subsection{The polymer model}
	
	A well known generic polymer model is used for the BBPs~\cite{kremer1990dynamics}. Within this model, 
	individual monomers interact via 6--12 Lennard--Jones (LJ) potential within a cut--off radius $r_{\rm c}$. 
	Unless stated otherwise $r_{\rm c} = 2^{1/6} \sigma$. The results are represented in the units of LJ energy $\epsilon$, LJ distance $\sigma$, mass of the 
	individual monomers $m$, and time $t_{\circ} = \sigma \sqrt{m/\epsilon}$.
	The numbers that are representative of hydrocarbons are
	$\epsilon = 30~{\rm meV} \simeq 1 k_{\rm B}T$, $\sigma = 0.5$ nm, and pressure $p_{\circ} = 40$ MPa~\cite{kremer1990dynamics}. 
	
	The bonded monomers interact with an additional finitely extensible nonlinear elastic (FENE) potential 
	$u_{\rm FENE}(r)= -0.5 k R_{\circ}^2 \ln \left[1 - \left({r}/{R_{\circ}}\right)^2\right]$,
	where $k = 30k_{\rm B}T/\sigma^2$ and $R_{\circ} = 1.5\sigma$. The effective bond length in this model is $\ell_{\rm b} \simeq 0.97\sigma$~\cite{kremer1990dynamics}. 
	
	Simulations are performed using the LAMMPS molecular dynamics package~\cite{thompson2022lammps,plimpton1995fast}. 
	The equations of motion are integrated using the velocity Verlet algorithm~\cite{verlet1967computer}. 
	The Langevin thermostat is employed to impose $T = 1\epsilon/k_{\rm B}$ with a damping coefficient $\gamma = 1.0t_{\circ}^{-1}$. 
	Note that the simulations are performed in different steps and thus the specific details will be presented wherever appropriate.
	
	\subsubsection{Tethered bottle--brushes}
	
	\begin{figure*}[ptb]
		\includegraphics[width=0.73\textwidth,angle=0]{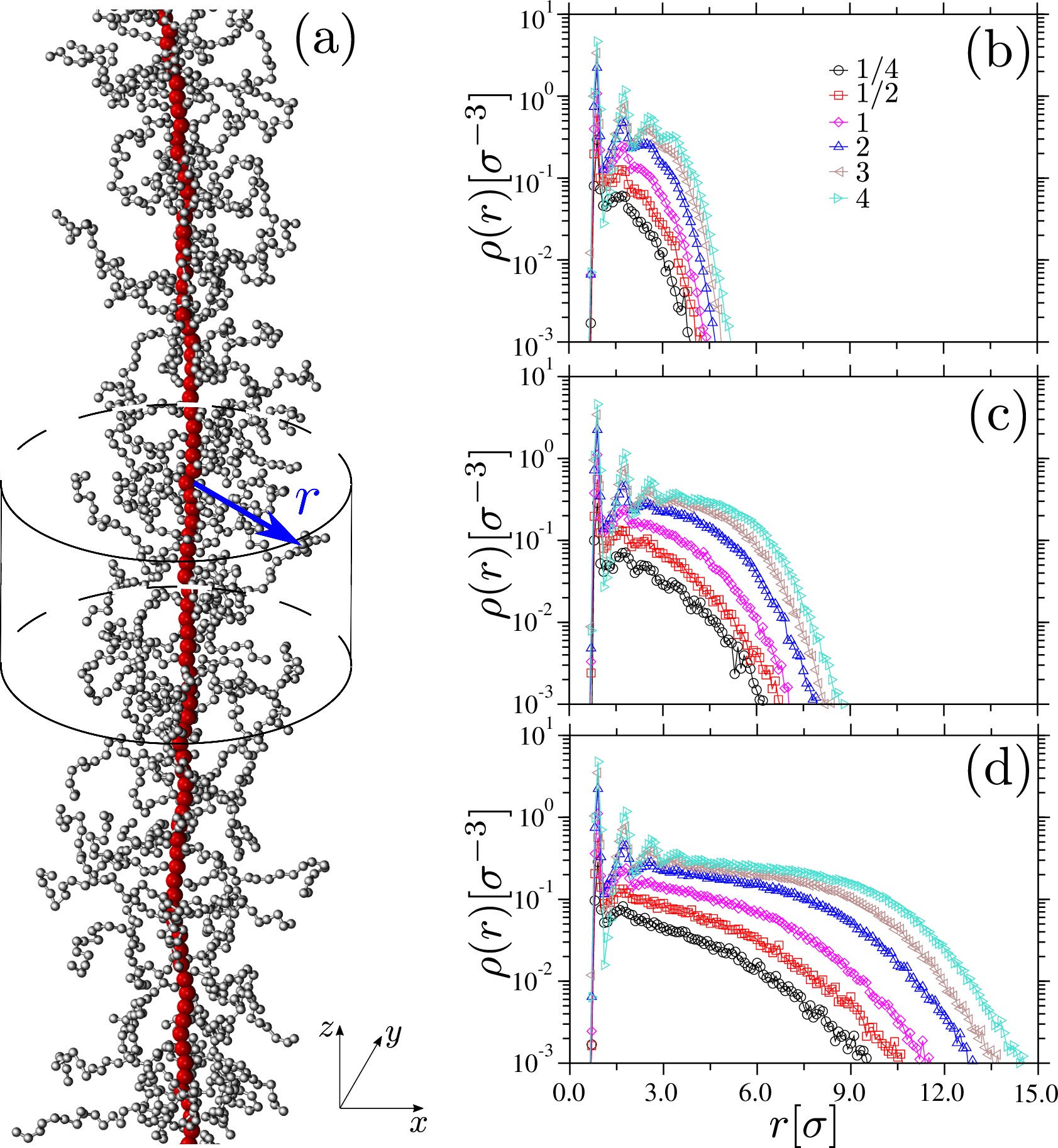}
		\caption{Part (a) shows a representative simulation snapshot of a tethered bottle--brush polymer (BBP). 
			This snapshot shows a section of a chain from the backbone length $N_{\rm b} = 500$, a side chain length $N_{\rm s} = 20$ 
			and a grafting density $\rho_{\rm g} = 1.0$, i.e., every backbone monomer is grafted with one side chain.
			The orientation of the chain is represented by the axis label in the right bottom corner. The thermal 
			transport coefficient $\kappa$ is calculated along the $z-$direction.
			The equilibrium monomer radial density profiles $\rho(r)$ are shown in parts (b--d). 
			$\rho(r)$ is shown for the different $\rho_{\rm g}$ and for $N_{\rm s} = 5$ (panel b), $N_{\rm s} = 10$ (panel c),
			and $N_{\rm s} = 20$ (panel d).
			Here, $\rho(r)$ is calculated in the cylindrical coordinate system, where the axis of a cylinder coincide 
			with the orientation of the BBP backbone, and $r$ is the radial distance from the backbone.
			\label{fig:BBPsnap}}
	\end{figure*}
	
	The tethered BBPs consist of $N_{\rm b} = 500$, $N_{\rm s} = 2-20$ and $\rho_{\rm g} = 1/256-4$. 
	A typical simulation snapshot of a BBP is shown in Fig.~\ref{fig:BBPsnap}(a). 
	The system size details for the tethered chains are listed in the Supplementary Table S1.
	The chain configurations are generated by constraining the first (index 1) and the last (index 500) backbone monomers. 
	These chains are first equilibrated under canonical simulation for a time $t = 10^5 t_{\circ}$ in their
	tethered positions. The time step is chosen as $\Delta t = 0.005t_{\circ}$. 
	The equilibrium side chain monomer density profiles $\rho(r)$ are shown in Figs.~\ref{fig:BBPsnap}(b--d) 
	for different $\rho_{\rm g}$ and $N_{\rm s}$.
	
	Periodic boundary conditions are employed in all three directions, however, the box size along the lateral directions 
	(i.e., x \& y directions) are taken as $2(d + r_{\rm c})$, where $d$ is the distance from the backbone at which 
	$\rho(r)$ drops to zero in Figs.~\ref{fig:BBPsnap}(b--d). This choice avoids a monomer to see its periodic image
	due to their lateral fluctuations. 
	
	\subsubsection{Bottle brush melts}
	
	For the melts, we have taken $N_{\rm b} = 50$ with $N_{\rm s} = 10$ and 20
	and $\rho_{\rm g} = 1/2$ and 1. The number of chains $N_{\rm c}$ in a melt is chosen such that the total 
	number of particles in a simulation box remains around $N \simeq 2.56 \times 10^5$, see the Supplementary Table S2. 
	Here, we have chosen a smaller value of $N_{\rm b}$ to ensure the reasonable sample equilibration, while
	avoiding the effects arising because of the chain entanglement. 
	Note also that the specific $N$ and $N_{\rm b} = 50$ are chosen to be consistent with the system
	size taken in an earlier study of two of us in Ref.~\cite{DMPRM21}.
	
	
	A melt configuration is generated by randomly placing $N_{\rm c}$ polymer chains in a cubic box at a monomer number density of $\rho_{\rm N} \simeq 0.85\sigma^{-3}$~\cite{kremer1990dynamics}.
	The melt samples are first equilibrated in the canonical ensemble for $t_{\rm equil} \simeq 4 \times 10^6 t_{\circ}$ with $\Delta t = 0.01t_{\circ}$. 
	The corresponding mean--squared displacements C$(t)$ are shown in the Supplementary Fig.~S1 and in the Supplementary Section~S2. 
	It can be appreciated that the total simulation time is at least an order of magnitude larger than 
	the typical relaxation times of all BBPs and thus attaining well--equilibrated state, see the Supplementary Fig. S1.
	
	For the $\kappa$ calculations, we have imposed an attractive LJ interaction between the non--bonded monomers with $r_{\rm c} = 2.5\sigma$. 
	Here, two different values are chosen to mimic the monomer--monomer interactions between the side chains, 
	namely $\epsilon_{\rm ss} = 1.0\epsilon$ (referred as the default system, as in an earlier work~\cite{DMPRM21}) 
	and $\epsilon_{\rm ss} = 1.2\epsilon$ (referred as the modified system). These systems are further equilibrated for $t_{\rm equil} \simeq 4 \times 10^5 t_{\circ}$ 
	with $\Delta t = 0.01t_{\circ}$ under the canonical ensemble. These systems are then density equilibrated at a
	bulk pressure of $p = 0 \epsilon/\sigma^3$ for $t_{\rm press} \simeq 4 \times 10^5 t_{\circ}$ with $\Delta t = 0.01t_{\circ}$.
	Pressure is imposed using the Nose-Hoover barostat~\cite{martyna1994constant,Benedict}.
	
	\subsection{$\kappa$ calculations}
	
	\subsubsection{Non--equilibrium approach--to--equilibrium method}
	\label{ss:apeqm}
	
	To calculate the thermal transport coefficient of full tethered chains $\kappa_{\rm full}$, we have 
	employed a non--equilibrium approach--to--equilibrium method~\cite{Lampin2013}. 
	Within this method, the length of a BBP $L_z$ is divided into two parts along the direction of heat flow (see Fig.~\ref{fig:BBPsnap}(a)).
	At the first step, 50 middle backbone monomers (including their side chains) are thermalized at $T_{\rm high} = 4\epsilon/k_{\rm B}$ 
	and the remaining monomers are kept at $T_{\rm low} = 0.67\epsilon/k_{\rm B}$, which ensures that 
	$\sum_{i = 1}^{N_{\rm sc}}T_i/N_{\rm sc} = 1.0 \epsilon/k_{\rm B}$ and $N_{\rm sc}$ 
	is the total number of particles in a tethered BBP. 
	This thermalization is performed for a total time of $t_{\rm therm}= 5 \times 10^5t_{\circ}$ with a time step of $\Delta t = 0.005t_{\circ}$. 
	Subsequently, the thermostat is switched off and $\Delta T (t) = T_{\rm high} - T_{\rm low}$ is allowed to
	relax under microcanonical ensemble for $t_{\rm relax}= 10^3t_{\circ}$ with $\Delta t = 0.001t_{\circ}$. 
	Following Ref.~\cite{Lampin2013}, bi--exponential relaxation $\Delta T (t) = c_1 {\rm exp}(-t/\tau) + c_2 {\rm exp}(-t/\tau_{z})$ 
	is used to obtain the time constant of the energy flow along the $z-$direction $\tau_z$. $\tau$, $c_1$ and $c_2$ are 
	the longitudinal energy transfer rate, and two constant pre--factors, respectively. 
	$\kappa_{\rm full}$ is then calculated using,
	\begin{equation}
		\kappa_{\rm full} = \frac{1}{4\pi^2} \frac{c L_z}{A \tau_z}.
		\label{apteq}
	\end{equation}
	Here, heat capacity is estimated using the classical Dulong--Petit limit $c = 3N_{\rm sc}k_{\rm B}$, $L_z = 499\ell_{\rm b}$, 
	and $A=\pi r^2$ is the cross--section area and the effective radius $r$ is estimated from $\rho(r)$ in Fig.~\ref{fig:BBPsnap}(b--d),
	i.e., when $\rho(r)$ drops to about 50\% of its maximum value~\cite{bhardwaj2021thermal}.
	See the Supplementary Table S3 for more details.
	
	\subsubsection{Equilibrium Kubo--Green method}
	\label{ss:kg}
	
	The Kubo-Green method~\cite{KuboG}, where $\kappa$ is estimated using,
	\begin{equation}
		\kappa=\frac{v}{Dk_{{\rm B}}T^{2}}\int_0^{\mathcal{T}} \left<{\bf J}(t)\cdot{\bf J}(0)\right> {\rm d}{t}.
		\label{eq:kappa}
	\end{equation}
	The heat flux auto--correlation function $\langle{\bf J}(t) \cdot {\bf J}(0)\rangle$ 
	is calculated in the microcanonical ensemble. Here, $v$ is the system volume, $D$ is the dimensionality, 
	and $\mathcal T$ is the range of integration that in our case is taken to be at least one order of magnitude larger than the typical de--correlation time.
	This method is used to calculate the backbone thermal transport coefficient $\kappa_{\rm BB}$ and for the melts $\kappa_{\rm melt}$.
	
	To calculate $\kappa_{\rm BB}$ we have used,
	\begin{equation}
		\kappa_{\rm BB}=\frac{v_{\rm BB}}{k_{{\rm B}}T^{2}}\int_0^{\mathcal{T}_{\rm BB}} \langle {J_{z}}(t) \cdot {J_{z}}(0) \rangle {\rm d}{t},
		\label{eq:kappabb}
	\end{equation}
	where volume of the backbone is estimated as $v_{\rm BB} = \pi \sigma^3 N_{\rm b}/6$~\cite{bhardwaj2021thermal} and $\mathcal{T}_{\rm BB} \simeq 100t_{\circ}$
	{with $\Delta t = 0.005 t_{\circ}$}. $\kappa_{\rm BB}$ is averaged over 20 independent runs.
	
	For the calculation of thermal transport coefficient of the BBP melts $\kappa_{\rm melt}$, we used have Eq.~\ref{eq:kappa}. 
	In this case, $\mathcal T \simeq 100t_{\circ}$ {with $\Delta t = 0.005 t_{\circ}$} and $\kappa_{\rm melt}$ is average of five independent runs.
	
	\subsubsection{Some notes on the choice of methods for $\kappa$ calculations}
	
	We have chosen two different methods for $\kappa$ calculations because of their individual advantages. 
	Furthermore, since both methods reproduce the same trends, we believe to be in the right method choices.
	
	The non--equilibrium methods usually suffer from length effects, especially in the quasi one--dimensional systems,
	because of the boundary scattering and hence leads to a smaller estimates of $\kappa$ that depend on $L_{z}$.
	Additionally, the approach--to--equilibrium is a relatively easy and computationally efficient method. 
	
	For the component--wise $\kappa$ and also melts with relatively smaller system sizes, the Kubo--Green method may be more
	suitable because it produces $\kappa$ equivalent to their values in the asymptotic limit. 
	Here, however, the Kubo--Green method usually requires very careful sampling of the heat 
	flux auto--correlation function and possible averaging over several independent runs. 
	
	All $\kappa$ data presented here are normalized relative to the thermal transport coefficient 
	of a linear chain (without the side chains) $\kappa_{\rm linear}$. Here, we have used 
	three estimates of $\kappa_{\rm linear}$ calculated for three different systems and two different methods. 
	Fig.~\ref{fig:kappa_Ns_full} uses $\kappa_{\rm linear} = 141 \pm 17 k_{\rm B}/t_{\circ}\sigma$ calculated 
	using Eq.~\ref{apteq}. For Figs.~\ref{fig:kappa_Ns} \& \ref{fig:kappa_mass}, $\kappa_{\rm linear} = 830 \pm 50 k_{\rm B}/t_{\circ}\sigma$ 
	calculated using Eq.~\ref{eq:kappabb}. For Fig.~\ref{fig:kappa_melt}, we have used $\kappa_{\rm linear} = 5.4 \pm 0.3 k_{\rm B}/t_{\circ}\sigma$ 
	taken from an earlier work of two of us~\cite{DMPRM21}.
	
	\section{Results and discussions}
	\label{sec:res}
	
	\subsection{Thermal conductivity of tethered chains}
	
	\subsubsection{Full bottle--brushes}
	
	\begin{figure}[ptb]
		\includegraphics[width=0.49\textwidth,angle=0]{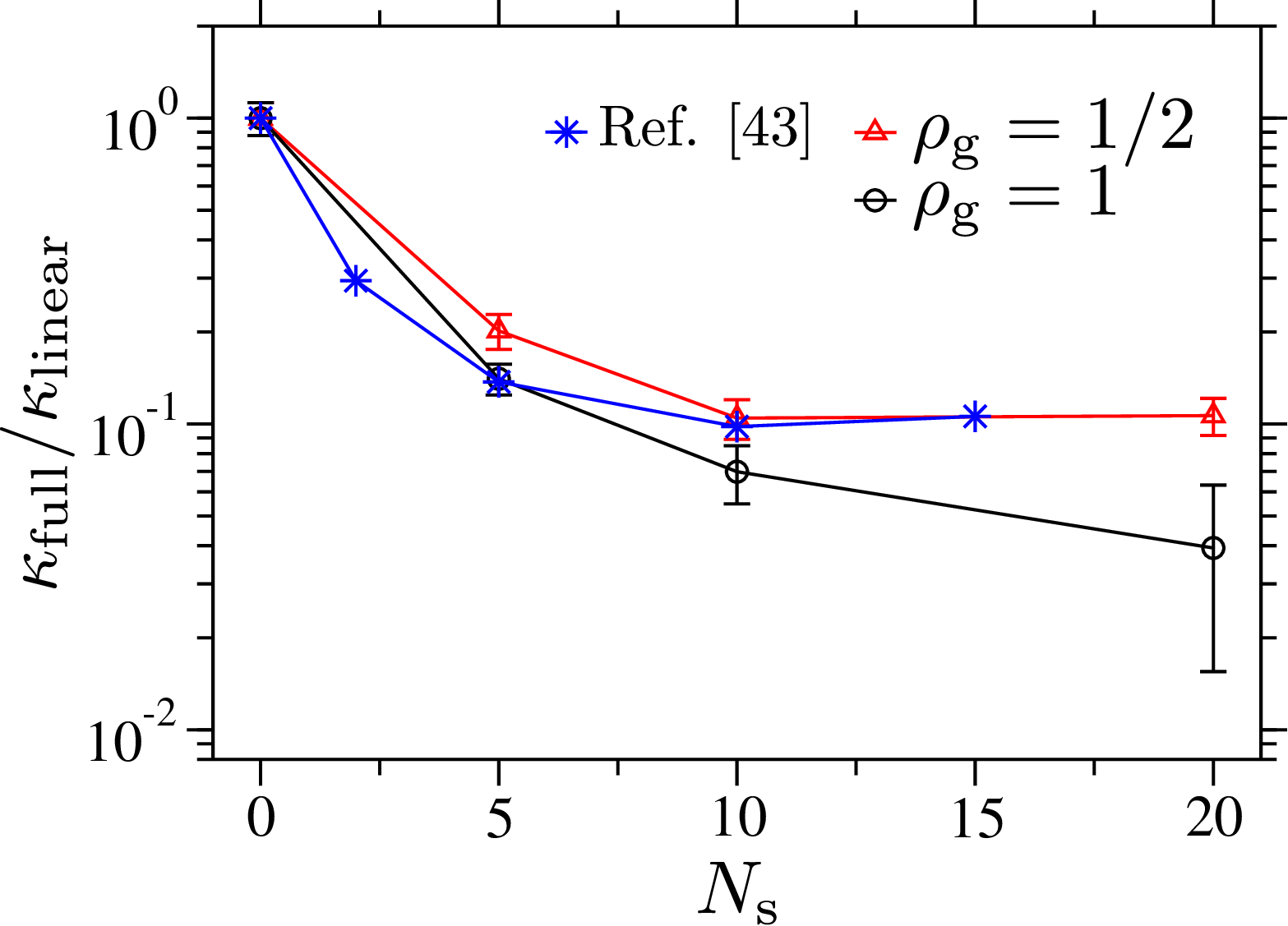}
		\caption{The normalized thermal transport coefficient of the full chain $\kappa_{\rm full}/\kappa_{\rm linear}$ 
			as a function of side chain length $N_{\rm s}$ for two different grafting densities $\rho_{\rm g}$.
			$\kappa_{\rm full}$ is normalized with the thermal transport coefficient of a single linear polymer without 
			the side chains $\kappa_{\rm linear} = 141 \pm 17 k_{\rm B}/t_{\circ}\sigma$. 
			The error bars are the standard deviation calculated from 20 independent simulation runs.
			For comparison, we have also included the all--atom reference data from Ref.~\cite{kappaHaoMaa}.
			\label{fig:kappa_Ns_full}}
	\end{figure}
	
	We begin by discussing the heat flow in full BBPs. In Fig.~\ref{fig:kappa_Ns_full}, we show $\kappa_{\rm full}/\kappa_{\rm linear}$ with changing $N_{\rm s}$ for 
	two different $\rho_{\rm g}$. Two distinct trends are clearly visible:
	
	\begin{itemize}
		
		\item $\kappa_{\rm full}$ decreases monotonically with increasing $N_{\rm s}$ for both $\rho_{\rm g}$. 
		This behavior is consistent with the trends observed in all--atom simulations of PNB grafted with PS~\cite{kappaHaoMaa}, where a reasonable quantitative agreement is observed. 
		
		\item $\kappa_{\rm full}$ decreases with increasing $\rho_{\rm g}$, i.e., going from the red to the black data set in Fig.~\ref{fig:kappa_Ns_full}. 
		This behavior is also consistent with the earlier results where $\kappa_{\rm full}$ was shown to decrease with increasing number of side chains along a backbone~\cite{kappaside18mat}. 
		
	\end{itemize}
	
	\begin{figure}[ptb]
		\includegraphics[width=0.49\textwidth,angle=0]{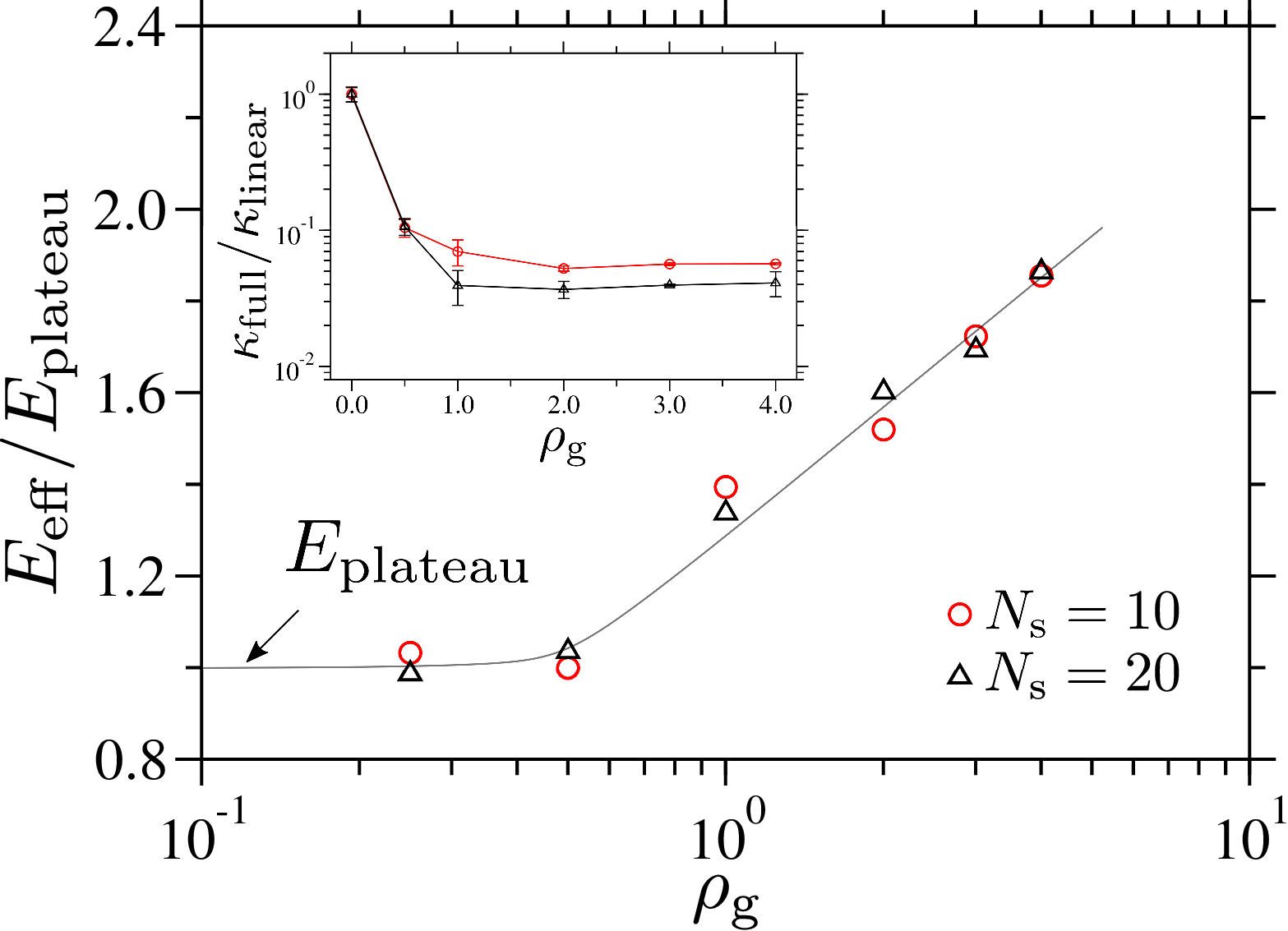}
		\caption{The main panel shows the normalized effective bending stiffness of a bottle brush $E_{\rm eff}$
			as a function of the grafting density $\rho_{\rm g}$. The data is normalized with the plateau value $E_{\rm plateau} = 6.7$
			for $\rho_{\rm g} \to 0$. Data is shown for two side chains lengths $N_{\rm s}$. In the inset we show the normalized thermal transport coefficient of full chain $\kappa_{\rm full}/\kappa_{\rm linear}$ as a function of $\rho_{\rm g}$.
			\label{fig:force}}
	\end{figure}
	
	While the data in Fig.~\ref{fig:kappa_Ns_full} show a nice correlation between the generic model and the corresponding all--atom data from the literature, 
	we will now focus on a more fundamental understanding. In this context, $\kappa_{\rm full}$ is governed by a delicate combination of different contributions:
	the bonded interactions along the backbone chain, the non--bonded contacts between the side chains, and the heat leakage pathways due to the side chains. 
	Here, a backbone chain consists of a periodic arrangement of monomers, where heat is carried by phonon propagation. 
	When defects appear along the backbone, phonons scatter and thus reduce $\kappa_{\rm full}$. 
	In BBPs, the defects arise from: (a) the flexural vibrations of the backbones itself and (b) the heat leakage due to the grafted side chains. 
	
	As of part (a), it is known that the backbone of a BBP becomes stiffer 
	with increasing $N_{\rm s}$ and/or by increasing $\rho_{\rm g}$, as estimated by $\ell_{\rm k}$~\cite{Sergei_1,binderJCP2011}. 
	The smaller the $\ell_{\rm k}$ value, the greater the number of kinks $N_{\rm kink}$ 
	along a backbone for a given $N_{\rm b}$. A defect kink along the backbone scatters phonon. 
	Indeed, it has been observed that increasing $N_{\rm kink}$ significantly reduces $\kappa$ of an isolated chain~\cite{GangChenJAP} and of a chain in a polymer brush~\cite{bhardwaj2021thermal}. 
	
	For a chain under the self--avoiding random walk configuration, $\ell_{\rm k}$ can be estimated 
	from the single chain structure factor~\cite{DesBook,DoiBook} or from the bond--bond auto--correlation function~\cite{binderJCP2011}. 
	However, for our tethered chains, we have first calculated the effective flexural stiffness by 
	using $E_{\rm eff} \simeq F/\delta$, where $F$ is the force required to keep the center monomer of a chain at a lateral displacement $\delta = 5\sigma$, 
	see the Supplementary Section~S3 and the Supplementary Fig.~S2 for more details. 
	In Fig.~\ref{fig:force}, we show the variation in $E_{\rm eff}$ with $\rho_{\rm g}$.
	It can be appreciated that $E_{\rm eff}$ increases with increasing $\rho_{\rm g}$, 
	which is expected given that the effective diameter of a BBP increases with 
	increasing $\rho_{\rm g}$ and the thicker cylinder give larger bending stiffness, 
	see Figs.~\ref{fig:BBPsnap}(b--d) and the Supplementary Table S3.
	
	Even when the data in Fig.~\ref{fig:force} reveal that the chains become stiffer with increasing $\rho_{\rm g}$, $\kappa_{\rm full}$ shows a qualitatively different 
	trend with $\rho_{\rm g}$, see the inset in Fig.~\ref{fig:force}.
	This highlight that the heat flow in BBP is rather non--trivial and that the side chains 
	play a more dominant role than just increasing the backbone flexural stiffness (scenario b, above). 
	We will again come back to this at a later stage of this draft.
	
	We note in passing that the flexural stiffness does not only increase with $\rho_{\rm g}$,
	rather it should (ideally) also increase with $N_{\rm s}$. Within the range of our choice of $N_{\rm s}$, 
	however, we do not observe any variation in $E_{\rm eff}$ with $N_{\rm s}$, 
	see the two data sets in the main panel of Fig.~\ref{fig:force}. This may be understood under two different conditions. 
	
	\begin{figure}[ptb]
		\includegraphics[width=0.49\textwidth,angle=0]{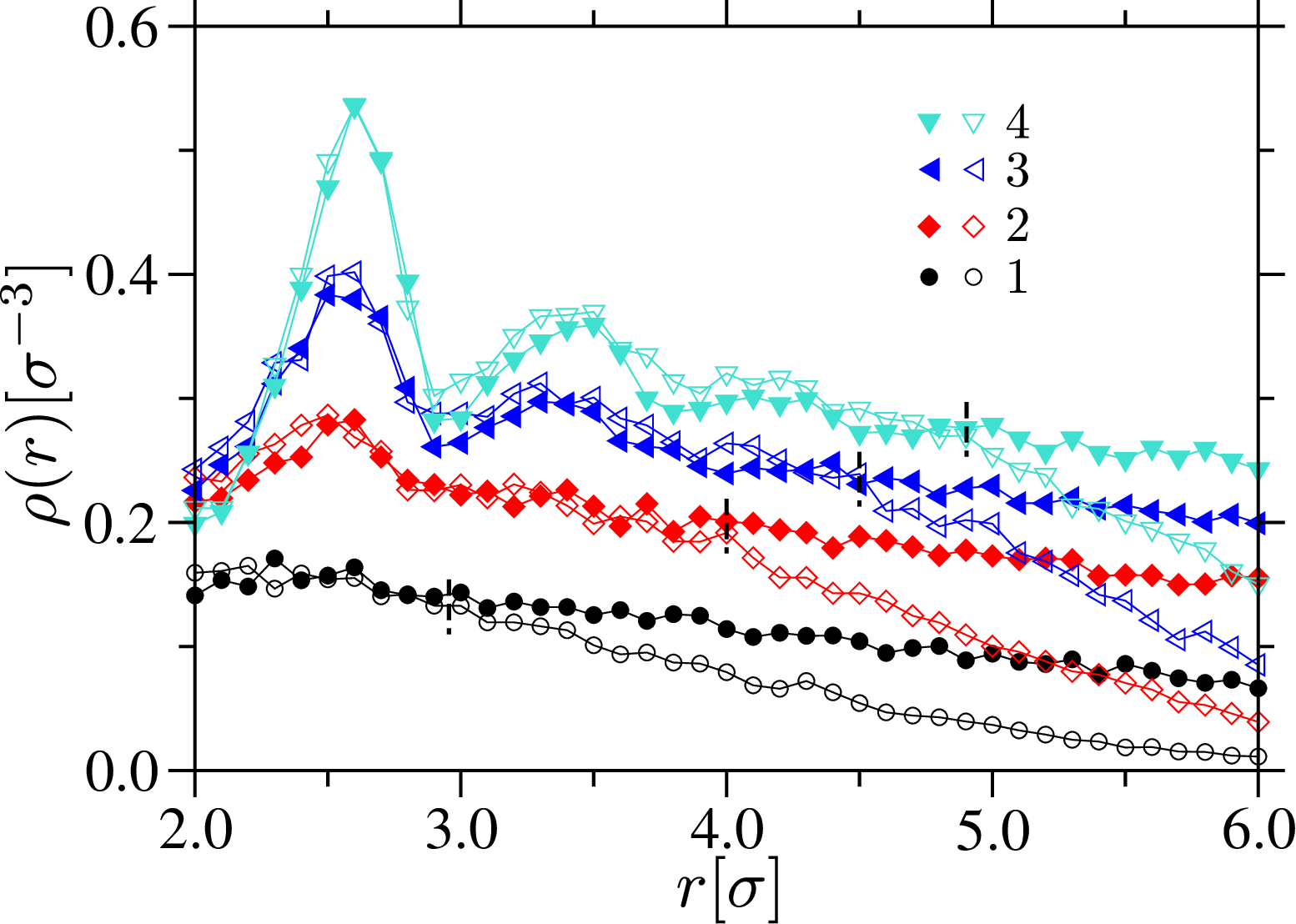}
		\caption{Same as Figs.~\ref{fig:BBPsnap}(c--d), however, only for the grafting density $\rho_{\rm g}$
			within the range 1--4. The data for the side chain length $N_{\rm s} = 10$ is shown by the empty symbols and
			the solid symbols are for $N_{\rm s} = 20$. Vertical lines in every set correspond to the $r$ values below 
			which both $N_{\rm s}$ have the same $\rho(r)$.
			\label{fig:rad_dens}}
	\end{figure}
	
	For the first condition, looking into $\rho(r)$ in Fig.~\ref{fig:rad_dens}, it can be appreciated that 
	$\rho(r)$ remains rather constant up to a distance (highlighted by the vertical black lines, which we refer to as $r^*$) 
	for both $N_{\rm s}$ and at a given $\rho_{\rm g}$.
	Furthermore, we assume that $E_{\rm eff}$ is majorly governed by the core, while the top layer of the a brush only act 
	as a soft surface that only contributes marginally to the total $E_{\rm eff}$. 
	Indeed, a previous study has shown that the stress on the top layers decreases rather rapidly (i.e., within a few particle diameters)~\cite{bhardwaj2021thermal}. 
	In this study, we have also attempted to calculate stress in BBPs, which, however, suffered from a poor signal to 
	noise ratio because of the short $N_{\rm s}$, and thus we abstain from this calculation.
	
	For the second condition, due to the short $N_{\rm s}$ values, chains also readjust upon lateral deformation and thus only
	contribute marginally to $E_{\rm eff}$. We expect the effect of $N_{\rm s}$ to be more significant on $E_{\rm eff}$ in the case 
	of longer side chains, especially for the grafting much larger than the critical grafting density for a $N_{\rm s}$.
	
	\subsubsection{Only backbone of the bottle--brushes}
	
	\begin{figure}[ptb]
		\includegraphics[width=0.43\textwidth,angle=0]{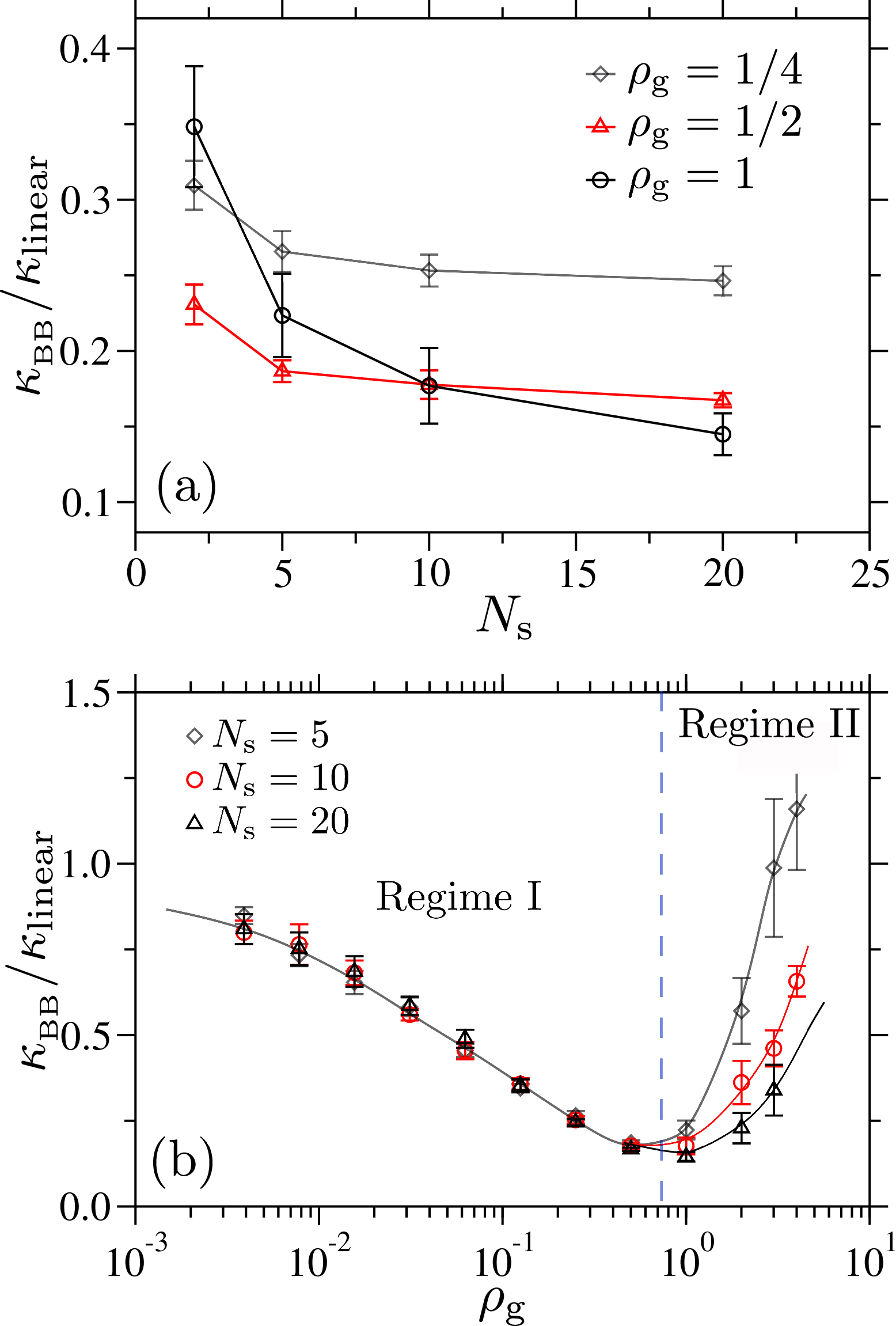}
		\caption{Part (a) shows the normalized thermal transport coefficient $\kappa_{\rm BB}/\kappa_{\rm linear}$ for the backbone only (without incorporating the side chains in $\kappa-$calculations) as a function of
			side chain length $N_{\rm s}$ for different grafting densities $\rho_{\rm g}$. Part (b) is a re--plot of the data in part (a) where 
			$\kappa_{\rm BB}/\kappa_{\rm linear}$ is shown as a function of $\rho_{\rm g}$ for different $N_{\rm s}$. 
			The vertical dashed line gives an indication of the two regimes of the $\kappa_{\rm BB}/\kappa_{\rm linear}$ behavior. The error bars are the standard deviation calculated from 20 independent simulation runs.
			$\kappa_{\rm BB}/\kappa_{\rm linear}$ is normalized with the thermal transport coefficient of a single linear polymer without the side chains 
			$\kappa_{\rm linear} = 830 \pm 50 k_{\rm B}/t_{\circ}\sigma$.
			\label{fig:kappa_Ns}}
	\end{figure}
	
	In this section, we will now focus on the backbone contributions to $\kappa_{\rm full}$. 
	For this purpose, we show $\kappa_{\rm BB}/\kappa_{\rm linear}$ as a function of $N_{\rm s}$ in Fig.~\ref{fig:kappa_Ns}(a),
	where $\kappa_{\rm BB}$ is the thermal transport coefficient only for the backbone monomers without incorporating the side chains into the calculations.
	$\kappa_{\rm BB}$ also decreases with increasing $N_{\rm s}$, 
	a similar trend as $\kappa_{\rm full}$ shown in Fig.~\ref{fig:kappa_Ns_full}. 
	A closer look at the data sets in Fig.~\ref{fig:kappa_Ns}(a) reveal a weak non--monotonic trend with increasing $\rho_{\rm g}$
	for a given $N_{\rm s}$. We have, therefore, re--plotted the data from Fig.~\ref{fig:kappa_Ns}(a) in Fig.~\ref{fig:kappa_Ns}(b), where 
	the variation in $\kappa_{\rm BB}$ with $\rho_{\rm g}$ is shown. To investigate the extent of backbone stiffening due to grafting, we have included additional data when $\rho_{\rm g}$ is increased such that every monomer is grafted with more than one side chain. It can be appreciated that $\kappa_{\rm BB}$ indeed increases when $\rho_{\rm g} \geq 1$ and thus highlight that the backbone stiffness is starting to contribute significantly to $\kappa_{\rm BB}$, while the heat leakage by the side chains gets compensated to some degree. 
	
	Within the picture discussed above, we can now identify two distinct regimes: 
	(I) For $\rho_{\rm g} \le 1/2$, $E_{\rm eff}$ remains rather constant and $\kappa_{\rm BB}$ is majorly influenced by the side chains.
	(II) For $\rho_{\rm g} \ge 1$, the backbone stiffening is dominant that overcomes scattering due to the side chains.
	These two observed regimes also explain why $\kappa_{\rm full}$ first decreases rapidly for $\rho_{\rm g} \le 1$, while 
	it remains rather constant for $\rho_{\rm g} \ge 1$ (or even shows a weak signature of increase),
	see the inset in Fig.~\ref{fig:force}. This later behavior predominantly comes from an increased $\kappa_{\rm BB}$ and 
	also because of the increased heat flow between the non--bonded contacts of the side chains for $\rho_{\rm g} \ge 1$ (data not shown).
	
	\begin{figure}[ptb]
		\includegraphics[width=0.49\textwidth,angle=0]{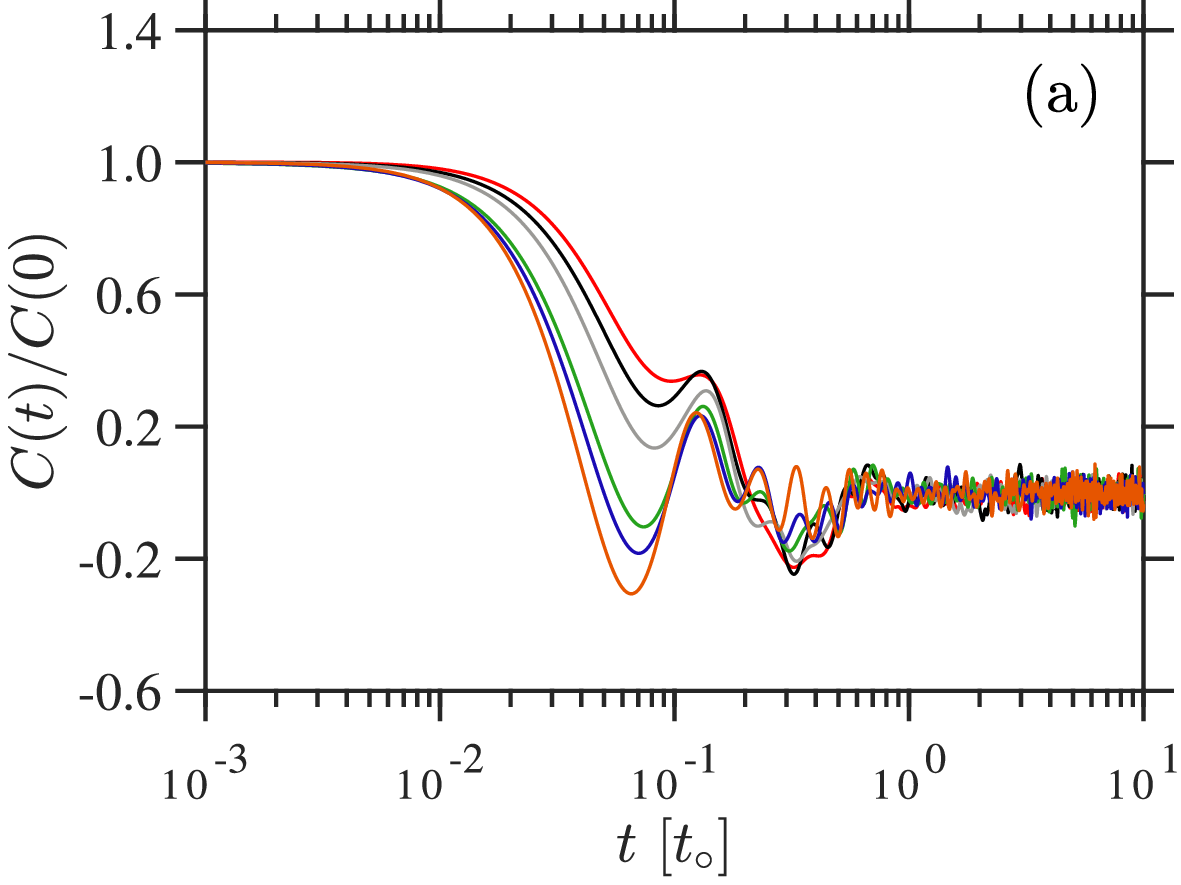}
		\includegraphics[width=0.49\textwidth,angle=0]{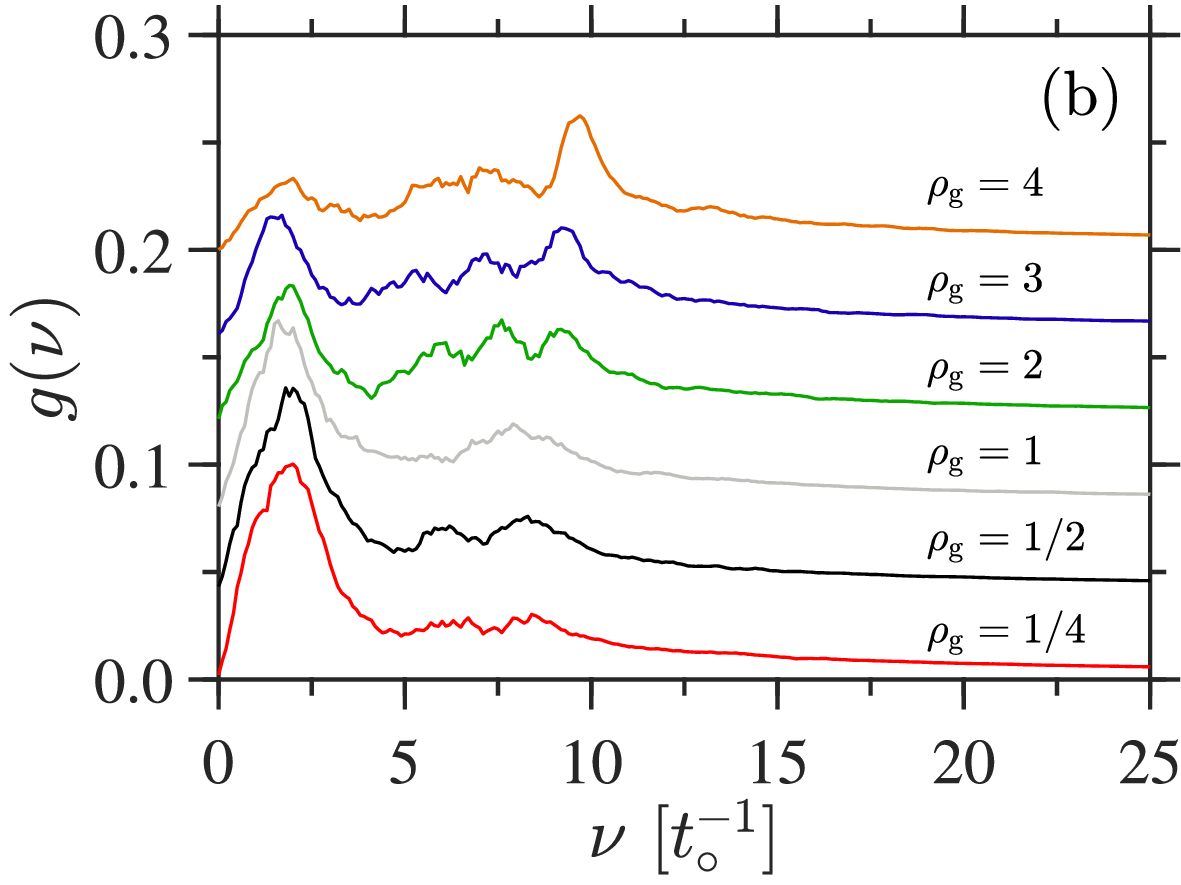}
		\caption{Part (a) shows the normalized mass--weighted velocity autocorrelation function $C(t)$, which is calculated for the backbone monomers only by using Eq.~\ref{eq:velcor}. Part (b) shows the vibrational density of states $g(\nu)$ calculated using Eq.~\ref{eq:vdos}. The data is presented for different grafting density $\rho_{\rm g}$ and for a side chain length $N_{\rm s} = 20$. For the clarity of presentation, individual $g(\nu)$ for different $\rho_{\rm g}$ are shifted by their additive y--offset. Colors are consistent between both panels.
			\label{fig:vdos}}
	\end{figure}
	
	{To illustrate the backbone stiffening and phonon scattering, we now calculate the vibrational density of states $g(\nu)$ with $\nu$ being the frequency. For this purpose, we first calculate the 
		mass--weighted velocity autocorrelation function,
		\begin{equation}
			C(t) = \sum_{i} m_i \langle {\overrightarrow v}(t) \cdot {\overrightarrow v}(0)\rangle.
			\label{eq:velcor}
		\end{equation}
		In Fig.~\ref{fig:vdos}(a) we show $C(t)/C(0)$ for a set of systems. The long lived
		fluctuations are clearly visible. 
		Here, the global $C(t)$ originate from the superposition of normal modes and thus its Fourier
		transform allows to compute $g(\nu)$ using~\cite{Horbach1999JPCB,martin21prm},
		\begin{equation}
			g(\nu) = \frac {1}{A}\int_{0}^{\infty} \cos(2\pi \nu t) \frac{C(t)}{C(0)} {\rm d}t,
			\label{eq:vdos}
		\end{equation}
		where the pre--factor $A$ ensures $\int g(\nu) {\rm d}\nu = 1$. Fig.~\ref{fig:vdos}(b) shows $g(\nu)$ 
		for different $\rho_{\rm g}$. A few distinct features are clearly visible:
		
		\begin{itemize}
			
			\item $g(\nu)$ shift towards higher $\nu$ with $\rho_{\rm g}$. This shift is an indication of backbone stiffening with increasing $\rho_{\rm g}$.
			
			\item The most prominent shift is observed for $\rho_{\rm g} \geq 1$, which is also consistent the trend in Fig.~\ref{fig:force}. These observations further show a nice correlation between increased stiffness and the observed apparent increase in $\kappa_{\rm BB}$, see Fig.~\ref{fig:kappa_Ns}.
			
			\item The flexural vibrational peak around $\nu \simeq 7-10 t_{\circ}^{-1}$ (going from $\rho_{\rm g} = 1/4-4$) becomes sharper with increasing $\rho_{\rm g}$. In this context, it is known that the width of a peak in $g(\nu)$ is inversely proportional to the phonon life time~\cite{KittelBook}, thus gives rise to a higher $\kappa_{\rm BB}$.
			
		\end{itemize}
		In summary, the non--monotonic behavior in Fig.~\ref{fig:kappa_Ns}(b) is because of the two competing effects. The initial decrease in $\kappa_{\rm BB}$ for $\rho_{\rm g}\to 1$ is dominated by the scattering via the presence of the side chain, where where the flexural stiffness remain rather invariant, see Figs.~\ref{fig:force} \& \ref{fig:vdos}.
		The further increase in $\kappa_{\rm BB}$ for $\rho_{\rm BB} \geq 1$ is because of the increased backbone stiffening that also leads to an increase in phonon life time.
		Note also that the peak around $\nu \simeq 2.5 t_{\circ}^{-1}$ comes from the non--bonded interactions.}
	
	\subsubsection{Effect of side chain mass}
	
\begin{figure}[ptb]
		\includegraphics[width=0.49\textwidth,angle=0]{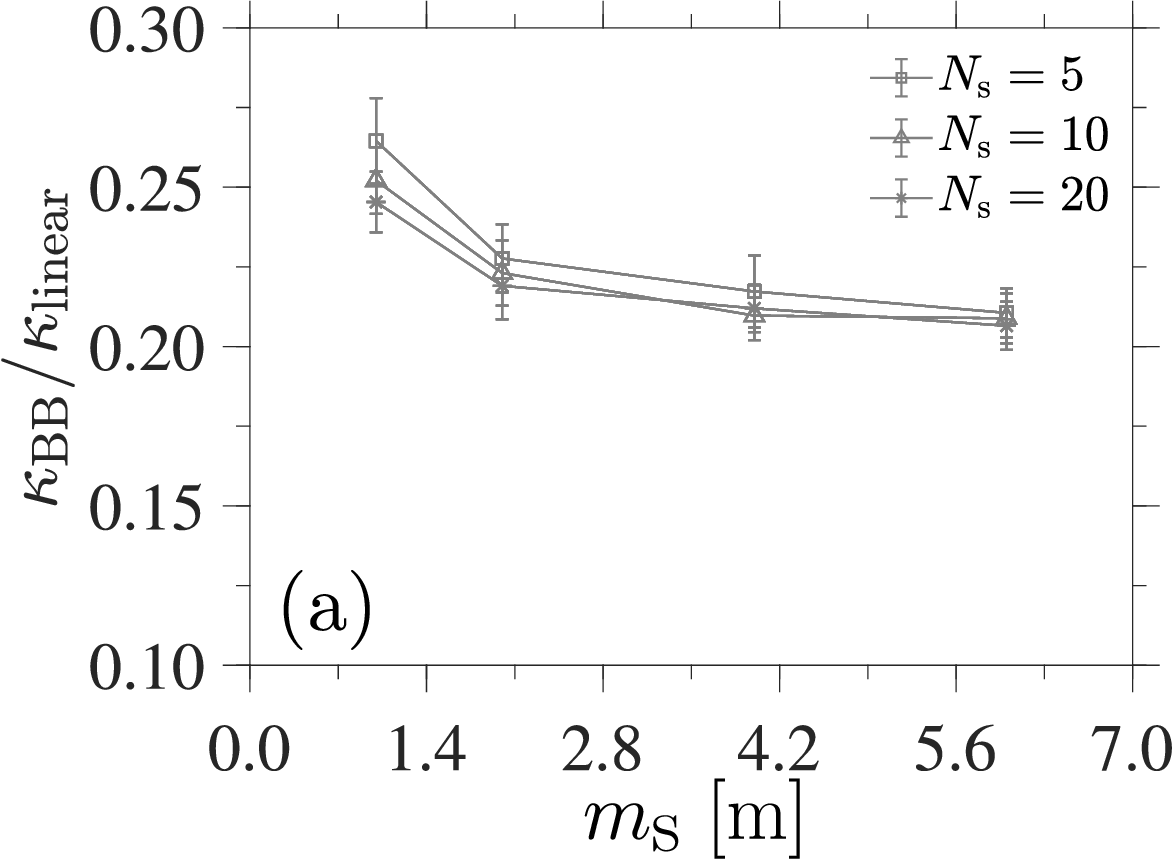}
		\includegraphics[width=0.49\textwidth,angle=0]{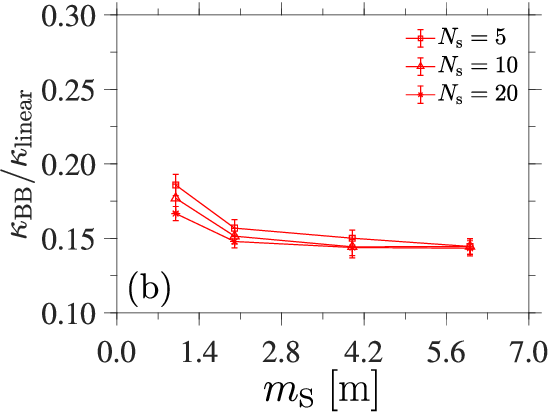}
		\includegraphics[width=0.49\textwidth,angle=0]{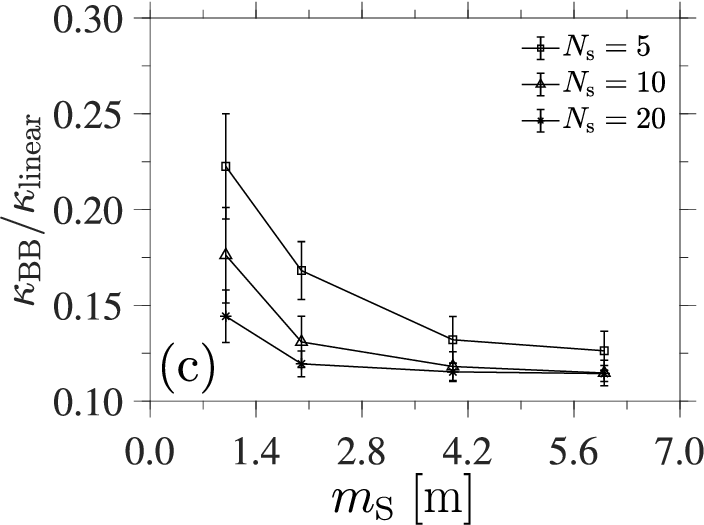}
		\caption{Same as Fig.~\ref{fig:kappa_Ns}(a), however, as a function of 
			monomer mass $m_{\rm s}$ of the side chains. The data is shown for three different 
			side chain lengths $N_{\rm s}$ and two different grafting densities $\rho_{\rm g}$. 
			Parts (a--c) correspond to $\rho_{\rm g} = 1/4$, 1/2, and 1, respectively.
			\label{fig:kappa_mass}}
	\end{figure}

	It is noteworthy that the presence of side chains has a more complex influence in dictating the overall heat flow, i.e., $N_{\rm s}$ directly influences $\kappa$. Here, however, increasing $N_{\rm s}$ also implies that the backbones are grafted with the bulkier side chains. Therefore, it is rather practical 
	to investigate: (1) What is the influence of the length $N_{\rm s}$? (2) What \% of contribution comes 
	directly from the increased the side chain mass? 
	To this end, we start by showing the change in $\kappa_{\rm BB}$ with the side chain monomer mass $m_{\rm s}$ in Fig.~\ref{fig:kappa_mass}. It can be appreciated that $\kappa_{\rm BB}$ decreases with increasing $m_{\rm s}$, where the most dominant effect is observed for $\rho_{\rm g} = 1$, see Fig.~\ref{fig:kappa_mass}(c). 
	This behavior is not surprising given that heat flow is directly proportional to the local vibrational frequencies and thus reduces with increasing $m_{\rm s}$ at a given $T$.
	
	\begin{table*}[ptb]
		\caption{A table listing the percentage change in the thermal transport coefficient of the backbone $\kappa_{\rm BB}$ for side chains with the same total mass but different side chain length $N_{\rm s}$. The data is compiled for three different grafting densities $\rho_{\rm g}$. The percentage difference is calculated with respect to the smaller $\kappa_{\rm BB}$ values. Here, $m_{\rm s}$ mass of the individual side chain monomer. $\kappa_{\rm BB}$ is normalized by the thermal transport coefficient of a linear tethered chain $\kappa_{\rm linear} = 830 \pm 50 k_{\rm B}/t_{\circ}\sigma$.}
		\centering
		\vspace{0.2in}
		\renewcommand{\arraystretch}{1.5}
		\begin{tabular}{ |c|c|c|c|c|c|c|c|}
			\cline{1-8}
			&&\multicolumn{2}{|c|}{$\rho_{\rm g}=1/4$} & \multicolumn{2}{|c|}{$\rho_{\rm g}=1/2$}  & \multicolumn{2}{|c|}{$\rho_{\rm g}=1$}   \\
			\cline{3-8}
			$N_{\rm s}$&$m_{\rm s}$[m]&$\kappa_{\rm BB}/\kappa_{\rm linear}$&Change[\%]&$\kappa_{\rm BB}/\kappa_{\rm linear} $&Change[\%]&$\kappa_{\rm BB}/\kappa_{\rm linear} $&Change[\%]\\
			
			
			\hline
			10&1&$  0.253 \pm   0.011$&& $0.178   \pm 0.009$&
			&$ 0.177   \pm 0.025$&\\
			
			&&&10.75&&12.80&&4.79\\
			
			5&2&$  0.229   \pm 0.011$&&$  0.158 \pm   0.006$&&$ 0.169  \pm  0.015$&\\
			\hline
			20&1&$  0.246   \pm 0.010$&&$  0.167  \pm  0.005$&& $ 0.145  \pm  0.014$&\\
			&&&10.50&&10.03&&10.26\\
			10&2&$  0.223  \pm  0.010$&&$0.152  \pm  0.005$& & $0.131 \pm  0.014$&\\
			\hline
			10&2&$  0.223  \pm  0.010$&&$0.152  \pm  0.005$& & $0.131 \pm  0.014$&\\
			&&&2.20&&0.91&&0.86\\
			5&4&$0.218 \pm 0.011$&&$ 0.151   \pm 0.005$&& $ 0.133  \pm  0.012$&\\
			\hline
			20&2&$   0.220  \pm  0.011$&&$ 0.149   \pm 0.004$&&$  0.120  \pm  0.007
			$&\\
			&&&4.46&&2.48&&1.12\\
			10&4&$   0.211  \pm  0.008$&&$   0.145  \pm  0.008$&&$0.119   \pm 0.009$&\\
			\hline
			
		\end{tabular} \label{tab:system}
	\end{table*}
	
	To identify the exact contribution due to the increased $m_{\rm s}$, we have compiled Table~\ref{tab:system}.
	Here, we show the percentage change in $\kappa_{\rm BB}$ by keeping the total mass $N_{\rm s} m_{\rm s}$ 
	of the individual side chains constant, but having different $N_{\rm s}$. It can be appreciated that the 
	increased $m_{\rm s}$ always has an additional contribution in knocking down $\kappa_{\rm BB}$. 
	For example, $\kappa_{\rm BB}$ changes by an additional 10--13\% for $m_{\rm s} \le 2$, while this 
	effect is only about a few \% for $m_{\rm s}> 2$. This is a direct consequence of the 
	behavior in Fig.~\ref{fig:kappa_mass}, where a rapid decrease in $\kappa_{\rm BB}$ is observed for 
	the smaller $m_{\rm s}$ and a weaker variation at higher $m_{\rm s}$. We, however, can not state why 
	exactly there is a significant difference between the two regimes, except the fact that mass 
	seemingly always has a greater effect on the heat flow than the exact $N_{\rm s}$. 
	
	We also want to briefly discuss a possible experimental system that could mimic the effect of mass difference, 
	while keeping $N_{\rm s}$ fixed. For example, when the short alkane chains are added as the side groups, such as in the case of the conjugated polymers~\cite{nancy22rev}, they may be replaced with polytetrafluoroethylene (PTFE). Note that one central difference between an alkane and a PTFE is that the hydrogen atoms are replaced with fluorines, hence effectively increasing the mass of a monomer by over a factor of three.
	
	\subsection{Thermal conductivity of bottle brush melts}
	
	\begin{figure}[ptb]
		\includegraphics[width=0.49\textwidth,angle=0]{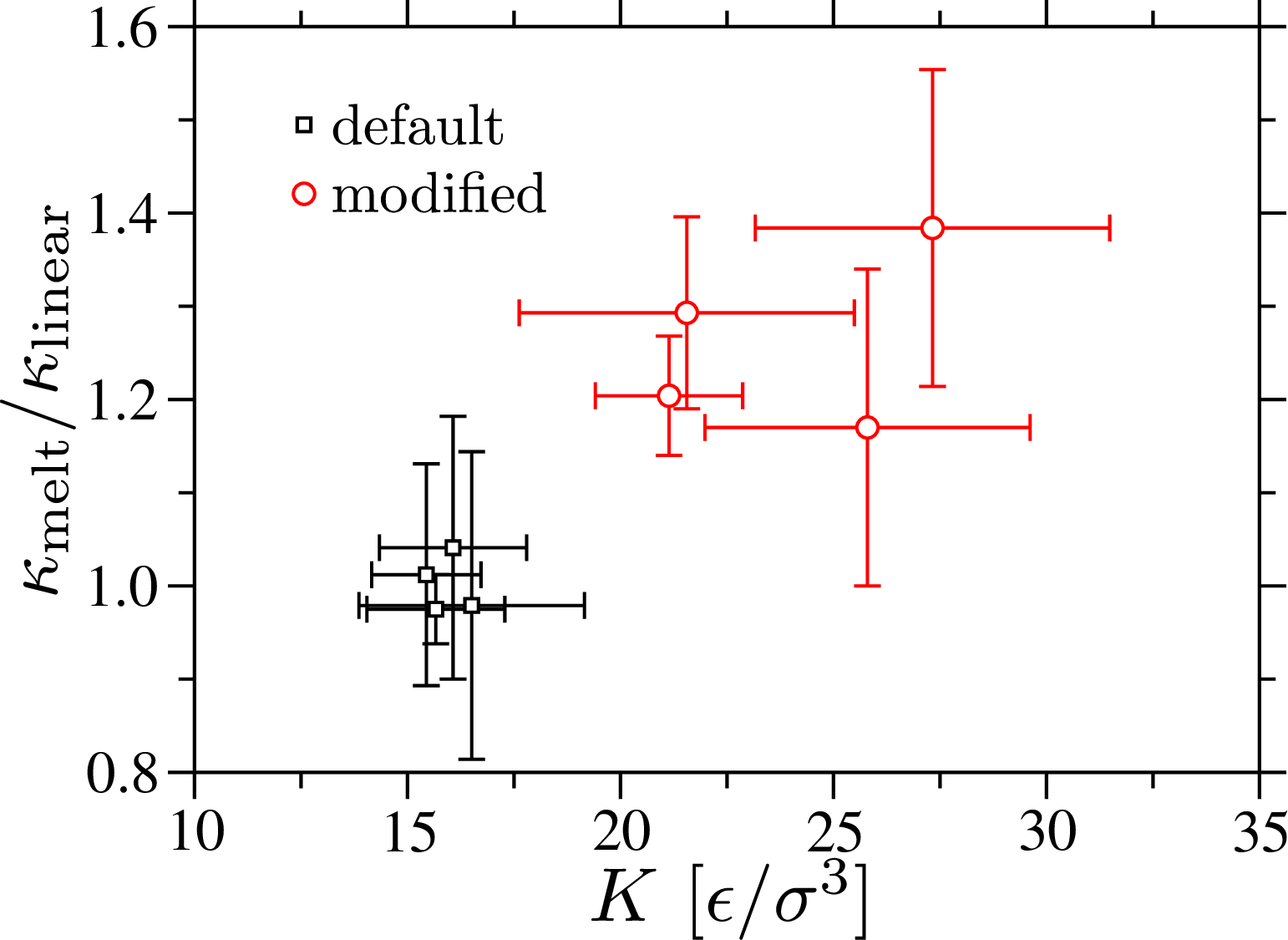}
		\caption{The normalized thermal transport coefficient of bottle--brush polymer melt $\kappa_{\rm melt}/\kappa_{\rm linear}$ as a function of the bulk modulus $K$. 
			The data is shown for two different interaction parameters between the monomers of the side chains, i.e., $\epsilon_{\rm ss} = 1.0\epsilon$ (default system)
			and $\epsilon_{\rm ss} = 1.2\epsilon$ (modified system). All other non--bonded monomer--monomer interaction is the same as the default system parameter.
			The thermal conductivity of the linear chain melt
			$\kappa_{\rm linear} = 5.4 \pm 0.3 k_{\rm B}/t_{\circ}\sigma$~\cite{DMPRM21}. 
			The error bars are the standard deviation calculated from five independent simulation runs.
			\label{fig:kappa_melt}}
	\end{figure}
	
	Lastly, we would like to briefly discuss heat flow in BBP melt. In Fig.~\ref{fig:kappa_melt}, we show the thermal transport coefficient of BBP melts $\kappa_{\rm melt}$ as a function of bulk modulus $K$, which is calculated using the volume fluctuation $K = k_{\rm B}T \left<v\right>/\left[\left<v^2\right> - \left<v\right>^2\right]$. It can be appreciated that the systems with the default interaction (black data set in Fig.~\ref{fig:kappa_melt}) have the same heat flow as the linear chain melt, i.e., $\kappa_{\rm melt} \simeq \kappa_{\rm linear}$, 
	which is independent of the exact BBP architecture and $N_{\rm s}$. This is, however, not surprising given that $\kappa$ in melts, consisting of flexible polymers, is predominantly dictated by the non--bonded interactions~\cite{Cahill16Mac,DMPRM21}. Only when the interaction between the side chain monomers is increased (see the red data set in Fig.~\ref{fig:kappa_melt}) can one observe an increase in $K$ and thus $\kappa_{\rm melt}$. In this context, it is important to highlight that there exists peptide--based BBP~\cite{weil20jacs} where side chains can have significantly larger interaction strengths, which may serve as a possible candidate in designing ``smart" commodity plastics with improved thermal properties.
	
	\section{Conclusions}
	\label{sec:conc}
	
	We have presented a molecular dynamics study of heat flow in tethered bottle--brush polymers (BBP) and BBP melts, 
	as quantified by the thermal transport coefficient $\kappa$. Our results show how the system parameters, such as the 
	grafted side chain length $N_{\rm s}$ and their density $\rho_{\rm g}$, control $\kappa$ of the tethered BBP. 
	In particular, we identify two different regimes in $\rho_{\rm g}$: for $\rho_{\rm g} < 1$ (weakly grafting regimes) 
	scattering due to the side chains dominate $\kappa$, while the backbone stiffening plays a dominant when 
	$\rho_{\rm g} \ge 1$ (highly grafting regime). Polymer architecture does not play a significant role in 
	dictating $\kappa$ in BBP melts, where the interactions between the side chains are a dominant factor in controlling $\kappa$. 
	As a broader perspective, our results establishes a structure--property relationship that may provide a 
	guiding tool in designing advanced soft materials with improved thermal properties.\\
	
	\noindent {\bf Supporting Information:} This file contains additional data presenting the details of system sizes, the cross--section of BBPs, the equilibration of BBP melts, and the stiffness calculations.
	
	\noindent {\bf Acknowledgement:} M.K.M. and M.K.S. thank National Supercomputing Mission (NSM) facility PARAM Sanganak at IIT Kanpur where most of the single-chain simulations were performed. M.K.S. also thanks funding support provided by IIT Kanpur under initiation grant scheme. D.M. thanks the ARC Sockeye facility where some of these simulations are performed. For D.M. this research was undertaken thanks, in part, to the Canada First Research Excellence Fund (CFREF), Quantum Materials and Future Technologies Program. 
	
	\noindent {\bf Data availability:} The scripts and the data associated with this research are available upon reasonable request from the corresponding author(s).
	
	\noindent {\bf Competing interests:} The authors declare no competing interests.
	\bibliographystyle{ieeetr}
	\bibliography{reference}

\begin{thebibliography}{10}

\bibitem{PolRevTT14}
A.~Henry, ``Thermal transport in polymers,'' {\em Annu. Rev. Heat Trans.},
  vol.~17, pp.~485--520, 2014.

\bibitem{Mueller20PPS}
M.~M\"uller, ``Process-directed self-assembly of copolymers: Results of and
  challenges for simulation studies,'' {\em Prog. Polym. Sci.}, vol.~101,
  p.~101198, 2020.

\bibitem{Mukherji20AR}
D.~Mukherji, C.~M. Marques, and K.~Kremer, ``Smart responsive polymers:
  Fundamentals and design principles,'' {\em Annu. Rev. Cond. Mat.}, vol.~11,
  pp.~271--299, 2020.

\bibitem{Keblinski20}
P.~Keblinski, ``Modeling of heat transport in polymers and their
  nanocomposites,'' {\em Handbook of Mater. Model.}, pp.~975--997, 2020.

\bibitem{PolRev2020Jie}
X.~Xu, J.~Zhou, and J.~Chen, ``Thermal transport in conductive polymer–based
  materials,'' {\em Adv. Funct. Mater.}, vol.~30, no.~8, p.~1904704, 2020.

\bibitem{nancy22rev}
N.~C. Forero-Martinez, K.-H. Lin, K.~Kremer, and D.~Andrienko, ``Virtual
  screening for organic solar cells and light emitting diodes,'' {\em Adv.
  Sci.}, vol.~9, no.~19, p.~2200825, 2022.

\bibitem{desiraju02}
G.~R. Desiraju, ``Hydrogen bridges in crystal engineering:interactions without
  borders,'' {\em Acc. Chem. Res.}, vol.~35, no.~7, pp.~565--573, 2002.

\bibitem{Pipe15NMat}
G.~Kim, D.~Lee, A.~Shanker, L.~Shao, M.~S. Kwon, Gidley, J.~Kim, and K.~P. Pipe
  {\em Nat. Mater.}, vol.~14, pp.~295--300, 2015.

\bibitem{Cahill16Mac}
X.~Xie, D.~Li, T.~Tsai, J.~Liu, P.~V. Braun, and D.~G. Cahill, ``Thermal
  conductivity, heat capacity, and elastic constants of water-soluble polymers
  and polymer blends,'' {\em Macromol.}, vol.~49, pp.~972--978, 2016.

\bibitem{weil20jacs}
C.~Chen, K.~Wunderlich, D.~Mukherji, K.~Koynov, A.~J. Heck, M.~Raabe, M.~Barz,
  G.~Fytas, K.~Kremer, D.~Y.~W. Ng, and T.~Weil, ``Precision anisotropic brush
  polymers by sequence controlled chemistry,'' {\em J. Am. Chem. Soc.},
  vol.~142, no.~3, pp.~1332--1340, 2020.

\bibitem{DMKKpolRev23}
D.~Mukherji and K.~Kremer, ``Smart polymers for soft materials: from solution
  processing to organic solids,'' {\em Polymers}, vol.~15, no.~15, p.~3229,
  2023.

\bibitem{review23}
B.~Mendrek, N.~Oleszko-Torbus, P.~Teper, and A.~Kowalczuk, ``Towards next
  generation polymer surfaces: Nano- and microlayers of star macromolecules and
  their design for applications in biology and medicine,'' {\em Prog. Polym.
  Sci.}, vol.~139, p.~101657, 2023.

\bibitem{halek1988relationship}
G.~W. Halek, ``Relationship between polymer structure and performance in food
  packaging applications,'' {\em ACS Symp. Ser.}, vol.~365, pp.~195--202, 1988.

\bibitem{jain2011biodegradable}
J.~Jain, W.~Y. Ayen, A.~J. Domb, and N.~Kumar, {\em Chapter 1: Biodegradable
  polymers in drug delivery}.
\newblock John Wiley \& Sons, Inc. New York, 2011.

\bibitem{maier2001polymers}
G.~Maier, ``Polymers for microelectronics,'' {\em Mater. Today}, vol.~4, no.~5,
  pp.~22--33, 2001.

\bibitem{Mukherji19PRM}
C.~Ruscher, J.~Rottler, C.~E. Boott, M.~J. MacLachlan, and D.~Mukherji,
  ``Elasticity and thermal transport of commodity plastics,'' {\em Phys. Rev.
  Mater.}, vol.~3, p.~125604, 2019.

\bibitem{pedot16}
W.~Quir{\'o}s-Solano, N.~Gaio, C.~Silvestri, G.~Pandraud, and P.~M. Sarro,
  ``Pedot: Pss: a conductive and flexible polymer for sensor integration in
  organ-on-chip platforms,'' {\em Procedia. Eng.}, vol.~168, pp.~1184--1187,
  2016.

\bibitem{shi2017tuning}
W.~Shi, Z.~Shuai, and D.~Wang, ``Tuning thermal transport in chain-oriented
  conducting polymers for enhanced thermoelectric efficiency: a computational
  study,'' {\em Adv. Funct. Mater.}, vol.~27, no.~40, p.~1702847, 2017.

\bibitem{tripathi2020optimization}
A.~Tripathi, Y.~Ko, M.~Kim, Y.~Lee, S.~Lee, J.~Park, Y.-W. Kwon, J.~Kwak, and
  H.~Y. Woo, ``Optimization of thermoelectric properties of polymers by
  incorporating oligoethylene glycol side chains and sequential solution doping
  with preannealing treatment,'' {\em Macromol.}, vol.~53, no.~16,
  pp.~7063--7072, 2020.

\bibitem{smith2016high}
M.~K. Smith, V.~Singh, K.~Kalaitzidou, and B.~A. Cola, ``High thermal and
  electrical conductivity of template fabricated p3ht/mwcnt composite
  nanofibers,'' {\em ACS Appl. Mater. Interfaces}, vol.~8, no.~23,
  pp.~14788--14794, 2016.

\bibitem{mcaninch2013characterization}
I.~M. McAninch, G.~R. Palmese, J.~L. Lenhart, and J.~J. La~Scala,
  ``Characterization of epoxies cured with bimodal blends of polyetheramines,''
  {\em J. Appl. Polym. Sci.}, vol.~130, no.~3, pp.~1621--1631, 2013.

\bibitem{elder2016nanovoid}
R.~M. Elder, D.~B. Knorr, J.~W. Andzelm, J.~L. Lenhart, and T.~W. Sirk,
  ``Nanovoid formation and mechanics: a comparison of poly (dicyclopentadiene)
  and epoxy networks from molecular dynamics simulations,'' {\em Soft Matter},
  vol.~12, no.~19, pp.~4418--4434, 2016.

\bibitem{DesBook}
J.~D. Cloizeaux and G.~Jannink, {\em Polymers in Solution: Their Modelling and
  Structure}.
\newblock Clarendon Press, 1990.

\bibitem{DoiBook}
M.~Doi and S.~F. Edwards, {\em The Theory of Polymer Dynamics}.
\newblock UK: Oxford Science Publications, 1986.

\bibitem{DGbook}
P.-G. de~Gennes, {\em Scaling Concepts in Polymer Physics}.
\newblock Cornell University Press, 1979.

\bibitem{MM21acsn}
S.~Gottlieb, L.~Pigard, Y.~K. Ryu, M.~Lorenzoni, L.~Evangelio,
  M.~Fernández-Regúlez, C.~D. Rawlings, M.~Spieser, F.~Perez-Murano,
  M.~Müller, and A.~W. Knoll, ``Thermal imaging of block copolymers with
  sub-10 nm resolution,'' {\em ACS Nano}, vol.~15, no.~5, pp.~9005--9016, 2021.

\bibitem{MM21mac}
L.~Pigard, D.~Mukherji, J.~Rottler, and M.~Müller {\em Macromol.}, vol.~54,
  no.~23, pp.~10969--10983, 2021.

\bibitem{wu22cms}
J.~Wu and D.~Mukherji, ``Comparison of all atom and united atom models for
  thermal transport calculations of amorphous polyethylene,'' {\em Comput.
  Mater. Sci.}, vol.~211, p.~111539, 2022.

\bibitem{Cahill90PRB}
D.~G. Cahill, S.~K. Watson, and R.~O. Pohl, ``Lower limit to the thermal
  conductivity of disordered crystals,'' {\em Phys. Rev. B}, vol.~46,
  pp.~6131--6140, 1990.

\bibitem{crist1996molecular}
B.~Crist and P.~G. Here{\~n}a, ``Molecular orbital studies of polyethylene
  deformation,'' {\em J Polym Sci B Polym Phys}, vol.~34, no.~3, pp.~449--457,
  1996.

\bibitem{GangChenNanoNature}
S.~Shen, A.~Henry, J.~Tong, R.~Zheng, and G.~Chen, ``Polyethylene nanofibres
  with very high thermal conductivities,'' {\em Nat. Nanotechnol.}, vol.~5,
  no.~4, pp.~251--255, 2010.

\bibitem{bhardwaj2021thermal}
A.~Bhardwaj, A.~S. Phani, A.~Nojeh, and D.~Mukherji, ``Thermal transport in
  molecular forests,'' {\em ACS Nano}, vol.~15, no.~1, pp.~1826--1832, 2021.

\bibitem{GangChenJAP}
X.~Duan, Z.~Li, J.~Liu, G.~Chen, and X.~Li, ``Roles of kink on the thermal
  transport in single polyethylene chains,'' {\em J. Appl. Phys.}, vol.~125,
  no.~16, p.~164303, 2019.

\bibitem{kappaside18mat}
C.~Huang, X.~Qian, and R.~Yang, ``Thermal conductivity of polymers and polymer
  nanocomposites,'' {\em Mater. Sci. Eng. R Rep.}, vol.~132, pp.~1--22, 2018.

\bibitem{superyellow}
S.~Schlisske, C.~Rosenauer, T.~Rödlmeier, K.~Giringer, J.~J. Michels,
  K.~Kremer, U.~Lemmer, S.~Morsbach, K.~C. Daoulas, and G.~Hernandez-Sosa,
  ``Ink formulation for printed organic electronics: Investigating effects of
  aggregation on structure and rheology of functional inks based on conjugated
  polymers in mixed solvents,'' {\em Adv. Mater. Technol.}, vol.~6, no.~2,
  p.~2000335, 2021.

\bibitem{Sergei_1}
J.~Paturej, S.~S. Sheiko, S.~Panyukov, and M.~Rubinstein, ``Molecular structure
  of bottlebrush polymers in melts,'' {\em Sci. Adv.}, vol.~2, no.~11,
  p.~e1601478, 2016.

\bibitem{binderJCP2011}
P.~E. Theodorakis, H.-P. Hsu, W.~Paul, and K.~Binder, ``{Computer simulation of
  bottle-brush polymers with flexible backbone: Good solvent versus theta
  solvent conditions},'' {\em J. Chem. Phys.}, vol.~135, p.~164903, 10 2011.

\bibitem{nanotube2016}
X.~Pang, Y.~He, J.~Jung, and Z.~Lin, ``1d nanocrystals with precisely
  controlled dimensions, compositions, and architectures,'' {\em Science},
  vol.~353, no.~6305, pp.~1268--1272, 2016.

\bibitem{lubrication}
X.~Banquy, J.~Burdynska, D.~W. Lee, K.~Matyjaszewski, and J.~Israelachvili,
  ``Bioinspired bottle-brush polymer exhibits low friction and amontons-like
  behavior,'' {\em J. Am. Chem. Soc.}, vol.~136, no.~17, pp.~6199--6202, 2014.

\bibitem{Sergei_2}
W.~F. Daniel, J.~Burdy{\'n}ska, M.~Vatankhah-Varnoosfaderani, K.~Matyjaszewski,
  J.~Paturej, M.~Rubinstein, A.~V. Dobrynin, and S.~S. Sheiko, ``Solvent-free,
  supersoft and superelastic bottlebrush melts and networks,'' {\em Nat.
  Mater.}, vol.~15, no.~2, pp.~183--189, 2016.

\bibitem{Sergei_4}
M.~R. Maw, A.~K. Tanas, E.~Dashtimoghadam, E.~A. Nikitina, D.~A. Ivanov, A.~V.
  Dobrynin, M.~Vatankhah-Varnosfaderani, and S.~S. Sheiko, ``Bottlebrush
  thermoplastic elastomers as hot-melt pressure-sensitive adhesives,'' {\em ACS
  Appl. Mater. Interfaces}, vol.~15, no.~35, pp.~41870--41879, 2023.

\bibitem{marquesThesis1989}
C.~M. Marques, {\em Les polym{\`e}res aux interfaces}.
\newblock PhD thesis, {\'E}diteur inconnu, 1989.

\bibitem{kappaHaoMaa}
H.~Ma and Z.~Tian, ``Effects of polymer topology and morphology on thermal
  transport: A molecular dynamics study of bottlebrush polymers,'' {\em Appl.
  Phys. Lett.}, vol.~110, no.~9, p.~091903, 2017.

\bibitem{kremer1990dynamics}
K.~Kremer and G.~S. Grest {\em J. Chem. Phys.}, vol.~92, no.~8, pp.~5057--5086,
  1990.

\bibitem{thompson2022lammps}
A.~P. Thompson, H.~M. Aktulga, R.~Berger, D.~S. Bolintineanu, W.~M. Brown,
  P.~S. Crozier, P.~J. {in 't Veld}, A.~Kohlmeyer, S.~G. Moore, T.~D. Nguyen,
  R.~Shan, M.~J. Stevens, J.~Tranchida, C.~Trott, and S.~J. Plimpton,
  ``Lammps-a flexible simulation tool for particle-based materials modeling at
  the atomic, meso, and continuum scales,'' {\em Comput. Phys. Commun.},
  vol.~271, p.~108171, 2022.

\bibitem{plimpton1995fast}
S.~Plimpton, ``Fast parallel algorithms for short-range molecular dynamics,''
  {\em J. Comput. Phys.}, vol.~117, no.~1, pp.~1--19, 1995.

\bibitem{verlet1967computer}
L.~Verlet, ``Computer" experiments" on classical fluids. i. thermodynamical
  properties of lennard-jones molecules,'' {\em Phys. Rev.}, vol.~159, no.~1,
  p.~98, 1967.

\bibitem{DMPRM21}
D.~Mukherji and M.~K. Singh, ``Tuning thermal transport in highly cross-linked
  polymers by bond-induced void engineering,'' {\em Phys. Rev. Mater.}, vol.~5,
  p.~025602, Feb 2021.

\bibitem{martyna1994constant}
G.~J. Martyna, D.~J. Tobias, and M.~L. Klein, ``Constant pressure molecular
  dynamics algorithms,'' {\em J. Chem. Phys.}, vol.~101, no.~5, pp.~4177--4189,
  1994.

\bibitem{Benedict}
B.~Leimkuhler and C.~Matthews, ``Molecular dynamics,'' {\em IJ. Interdiscip.
  Math.}, vol.~39, p.~443, 2015.

\bibitem{Lampin2013}
E.~Lampin, P.~L. Palla, P.-A. Francioso, and F.~Cleri, ``Thermal conductivity
  from approach-to-equilibrium molecular dynamics,'' {\em J. Appl. Phys.},
  vol.~114, no.~3, p.~033525, 2013.

\bibitem{KuboG}
R.~Zwanzig, ``Time-correlation functions and transport coefficients in
  statistical mechanics,'' {\em Annu. Rev. Phys. Chem.}, vol.~16, no.~1,
  pp.~67--102, 1965.

\bibitem{Horbach1999JPCB}
J.~Horbach, W.~Kob, and K.~Binder, ``Specific heat of amorphous silica within
  the harmonic approximation,'' {\em J. Phys. Chem. B}, vol.~103,
  pp.~4104--4108, May 1999.

\bibitem{martin21prm}
H.~Gao, T.~P.~W. Menzel, M.~H. M\"user, and D.~Mukherji, ``Comparing simulated
  specific heat of liquid polymers and oligomers to experiments,'' {\em Phys.
  Rev. Mater.}, vol.~5, p.~065605, Jun 2021.

\bibitem{KittelBook}
C.~Kittel, {\em Introduction to Solid State Physics, Eight Edition}.
\newblock John Wiley \& Sons, 2005.

\end{thebibliography}

	\begin{figure*}[ptb]
		\includegraphics[width=0.65\textwidth,angle=0]{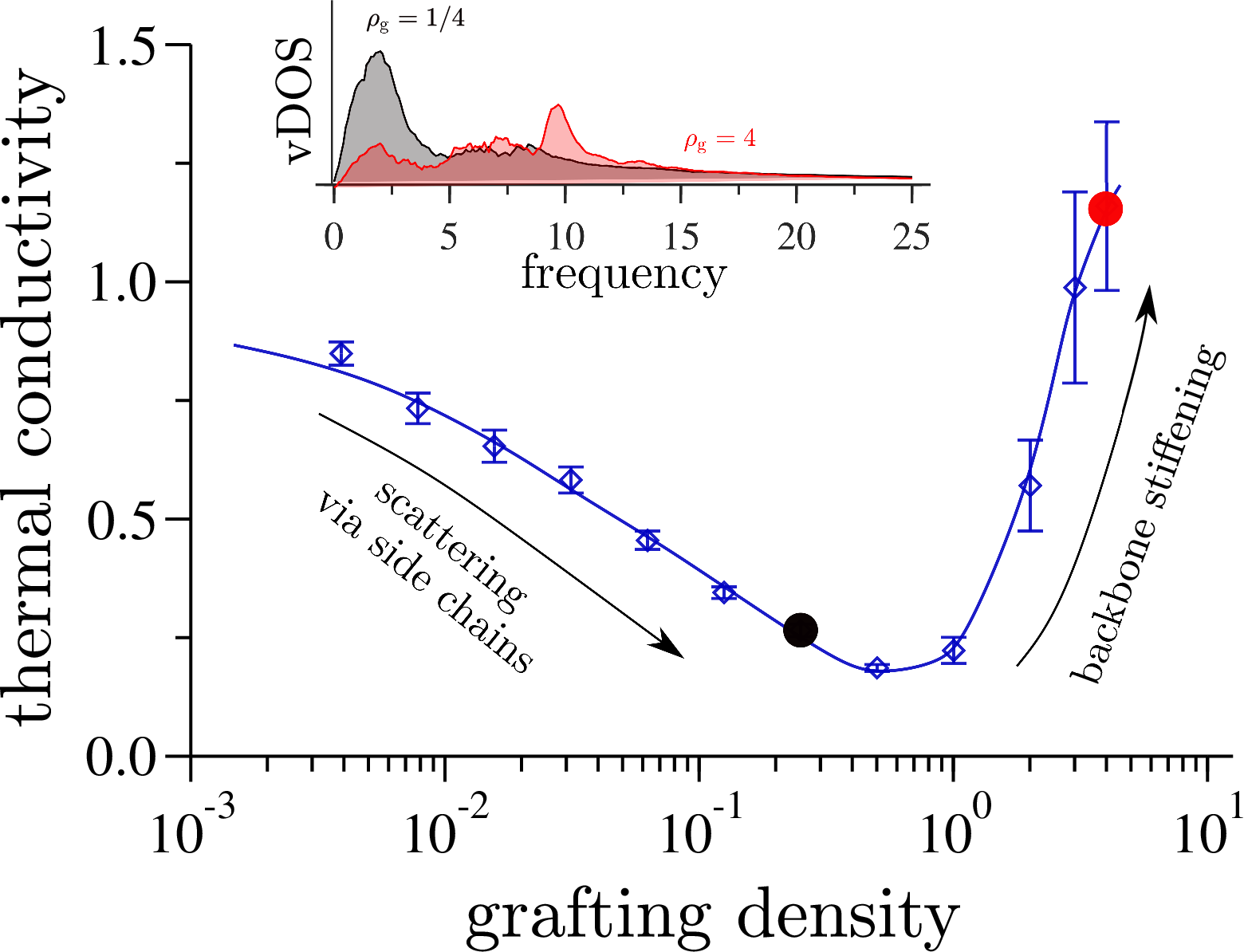}
	\end{figure*}

\end{document}